\documentclass[10pt,a4paper,twoside]{article}
\usepackage{epsfig}
\usepackage{baltlat6}
\usepackage{array}
\usepackage{here}
\usepackage{epstopdf}
\begin{document}
\ \ \vspace{0.5mm} \setcounter{page}{277} \vspace{8mm}


\titleb{ON THE GRAVITATIONAL STABILITY AND  MASS ESTIMATION OF STELLAR DISKS
}

\begin{authorl}
\authorb{A. Saburova}{1} and
\authorb{A. Zasov}{1}
\end{authorl}

\begin{addressl}
\addressb{1}{Sternberg Astronomical Institute of Moscow State University,
\\  Universitetskii pr. 13, Moscow, 119992, Russia;
saburovaann@gmail.com, zasov@sai.msu.ru}
\end{addressl}


\begin{summary}
We estimate the masses of disks of galaxies using the marginal gravitational stability
criterion and compare them with the photometrical disk mass evaluations. The comparison
reveals that the stellar disks of most of spiral galaxies we considered cannot be
substantially overheated (at least within several radial scalelengths) and are therefore
unlikely to have experienced a significant merging event in their history. However, for
substantial part of S0-type galaxies a stellar velocity dispersion is well in excess of
the gravitational stability threshold suggesting a major merger event in the past. For
four low surface brightness galaxies we found that the disk masses corresponding to the
marginal stability condition are significantly higher than it may be expected from their
brightness. Either their disks are dynamically overheated, or they contain a large amount
of non-luminous matter.
\end{summary}

\resthead{On the gravitational stability and mass estimation of stellar disks } {A.
Saburova, A. Zasov}
\sectionb{1}{INTRODUCTION}
A dynamical evolution of disks is the matter of hot debates, and the analysis of their kinematical characteristics plays a key role in the study of their
structure and history (see e.g. Zasov, Sil'chenko (2010) and references therein). By
kinematical characteristics we mean a rotation velocity and velocity dispersion of old
stars, making up the bulk of disk mass. In general, stellar velocity dispersion either can
reflect the velocity of turbulent motion of gas which gave the birth to the disk stars, or
can be the result of dynamical heating of disk caused by the internal or external reasons.
The minimal velocity dispersion of old stars at given $r$ is  constrained by disk local
marginal stability condition. Numerical simulations of models of initially weakly unstable
3D disks show the rapid transition of a disk into the marginally stable state, after which
the growth of velocity dispersion practically ceases (see e.g. Khoperskov et al. (2003)).

\sectionb{2}{THE LOCAL CRITERION OF GRAVITATIONAL STABILITY }
The constraints on the disk surface density distribution may be found from the radial
velocity dispersion, using the condition of gravitational stability of disk. The critical
value of radial velocity dispersion which makes a thin, collisionless isothermal disk
stable against the gravitational perturbations is usually written as:
$$ c_{cr}=Q_c\cdot c_{T} \approx Q_c\cdot \frac{3.4 \cdot G \cdot \sigma
_*}{\kappa},$$ where $Q_c$ is the stability parameter, $c_{T}$ is the Toomre critical radial velocity
dispersion,
 $\sigma _*$ is  the disk surface density, $\kappa$ is the epicyclical frequency.
  A finite disk thickness makes disk more stable, while
non-radial perturbations have the opposite effect.
The stability condition taking into account both of these effects can not be expressed in
analytic form. As numerical simulations of 2D and 3D disks show, the parameter $Q_c$ for a
wide range of $r$ lies in the interval 1.2 -- 2.5 (see e.g. Bottema (1993), Khoperskov et
al. (2003)).

Radial velocity dispersion can be estimated from the observed line-of-sight velocity
dispersion measured along the major axis: $c_{obs}(r) = (c_z \cdot cos^2 (i)+ c_{\phi}\cdot sin^2  (i) +
c_{r}sin^2(i))^{0.5}$, where $c_{obs} (r)$ is the line-of-sight velocity dispersion, $c_z$, $c_{\phi}$, $c_{r}$
are the vertical, azimuthal and radial components,  $i$ is the disk inclination. To
separate the components of the velocity dispersion one can introduce two additional
conditions: $ c_{r} = 2\Omega \cdot c_{\phi} /\kappa $ (Lindblad formula for the
epicyclical approximation) and $ c_z=m\cdot c_{r} $. Both the analysis of the available measurements of the galactic
disks velocity dispersion (Shapiro et al., 2003) and the results of numerical modeling
(see e.g. Zasov et al., 2008) show that in most cases $m\approx 0.4-0.7$.

By accepting these approximations and assuming that the disks are marginally stable, one
can put constraints on their surface densities $\sigma _*$. In general, when  marginal
stability condition does not hold, the resulting density and mass estimates can be treated as
 the upper limits.

\sectionb{3}{THE RESULTS OF DISK MASS ESTIMATIONS}
We used the marginal stability condition for galactic disks and the stellar velocity
dispersion data found in the literature for spiral and S0 galaxies to place the upper limits of the disk local
surface density $\sigma(2h)$ at the radial distance of about two radial scalelengths $r\approx2h$,
where the disk contribution to the observed velocity curve is maximal. Extrapolating these estimates, we constrained the total mass of the disks $M_d =
2\pi h^2\, \sigma(2h)\, e^2$ and compare these estimates to those based on the photometry
and color of stellar populations. We assumed $Q(2h)\approx 1.5$ (Khoperskov et al. (2003))
and the ratio $m=0.5$.  To compare the obtained disk masses with the photometric estimates
we calculated disk mass-to-light ratios in $B$ band $(M/L_B)_d$.  For Sa---S0 galaxies the contribution
 of bulge was taken into account (for more details see Zasov et al.
2011).

Diagrams ``$(M/L_B)_{d}$---$(B-V)_0$'' for stellar disks are shown in Fig. 1.  The
straight line reproduces model relation obtained by Bell, de Jong (2001) for stellar
systems with different present-day star formation rate (SFR) using the
bottom-light Salpeter initial mass function (IMF).
\begin{figure} [h!]
\includegraphics[width=6.0cm,keepaspectratio]{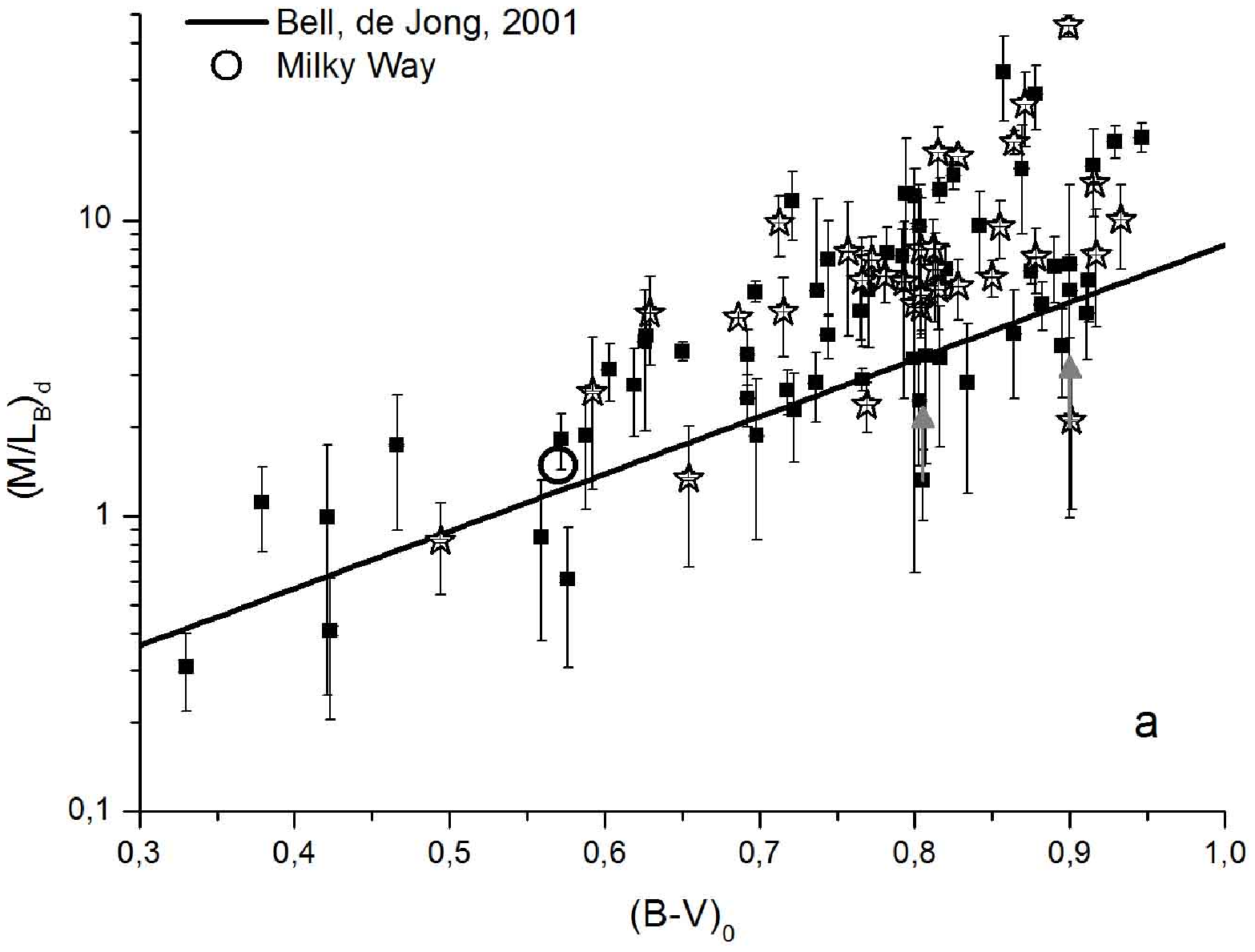}
\includegraphics[width=6.0cm,keepaspectratio]{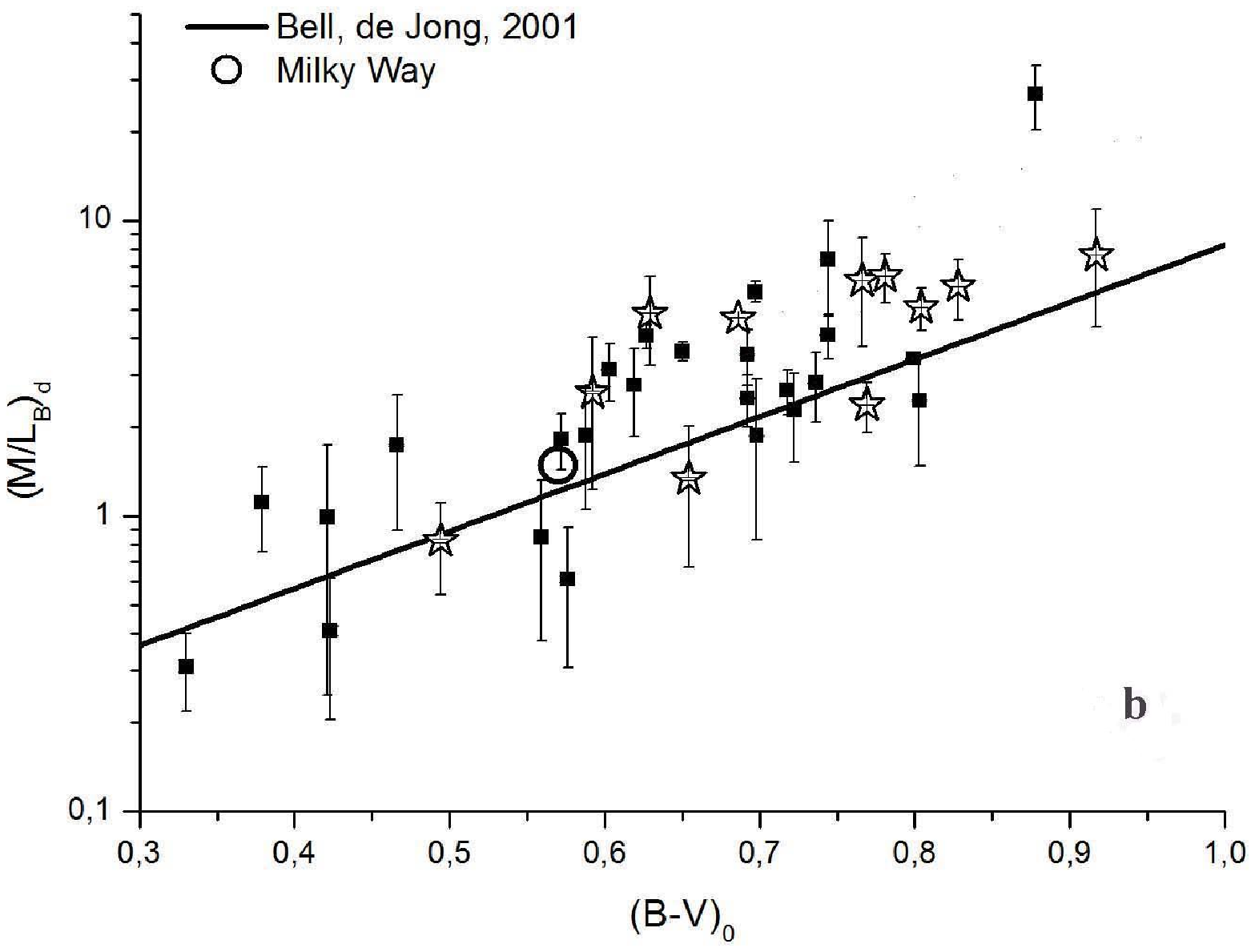}
\caption{Diagram ``$(M/L_B)_d$---$(B-V)_0$'', plotted for (a) the entire sample of
galaxies and (b) for the galaxies of types later than $S0/a$. Pair members are shown by asterisks. Straight line corresponds to photometric
model relations by Bell, de Jong (2001).}
\end{figure}
In Fig. 1a the entire sample of galaxies is shown; pair members
are marked by asterisks. The same diagram as in Fig. 1a, but after the
exclusion of S0/a---S0 galaxies is shown in Fig. 1b. Though the scatter of points in
these diagrams is large ($\sim 0.3~dex$), it is compatible with the errors of individual
mass estimates. Therefore there is a general agreement between
$(M/L_B)_{d}$ estimates based on marginal stability condition and those based on the
stellar population modeling. It is remarkable that most of the galaxies, which
significantly deviate from model relation, have a red color $(B-V)_0>0.7$. At least half of
these galaxies are above the straight line, that is most of them have disks with
a significant dynamical overheating. Note, that there are both paired and field galaxies
among the latters: it seems that only a strong gravitation perturbation can be a cause
of the disk dynamical overheating.

If to consider the disky galaxies as systems with marginally stable disks, one can plot
the relation $M_d$---$V_c$ (baryonic Tully-Fisher relation), connecting the most important
parameter of stellar disk (a mass) with the parameter determined mostly by massive halo
(a rotation velocity). This diagram is shown in Fig. 2. The relation obtained by McGaugh
(2005) for a large sample of spiral galaxies, where disk masses were estimated from their
luminosities and colors, is also shown by straight line. A comparison reveals the
absence of systematical difference between the two methods of mass estimation.
\begin{figure} [h!]
\includegraphics[width=6.0cm,keepaspectratio]{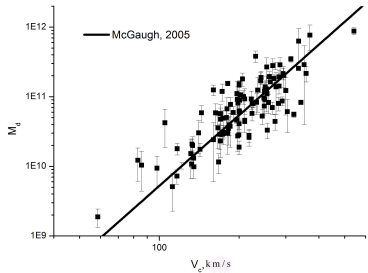}
\caption{Baryonic Tully-Fisher relation.
Straight line is taken from McGaugh (2005) for spiral galaxies where stellar masses were
found on the basis of the luminosities and colors of galactic disks.}
\end{figure}

A more detailed analysis was performed for the nearby galaxy M33. For this galaxy we applied
the modified marginal gravitational stability criterion taking into account the influence
of gas on the disk stability (see Saburova \& Zasov 2012). We used
the rotation curve of Corbelli (2003) and the radial profile of line-of-sight velocity
dispersion of the disk planetary nebulae obtained by Ciardullo et al. (2004). Fig. 3
demonstrates the radial profiles of local $M/L_K$, estimated for marginally stable disk. The K-band surface
brightness was taken from Regan, Vogel (1994).  The dotted line denotes $M/L_K$
profile taken from the photometric model of Bell, de Jong (2001) for the observed color
indices: $(B-V)$ for the inner part (Guidoni et al. 1981) and $(H-K)$ for the outer part
(Regan, Vogel 1994) of the disk. A thin dotted line in Fig. 3a includes the  correction for
internal dust extinction (Verley et
al. 2009).

\begin{figure}[h!]
\includegraphics[width=6.0cm,keepaspectratio]{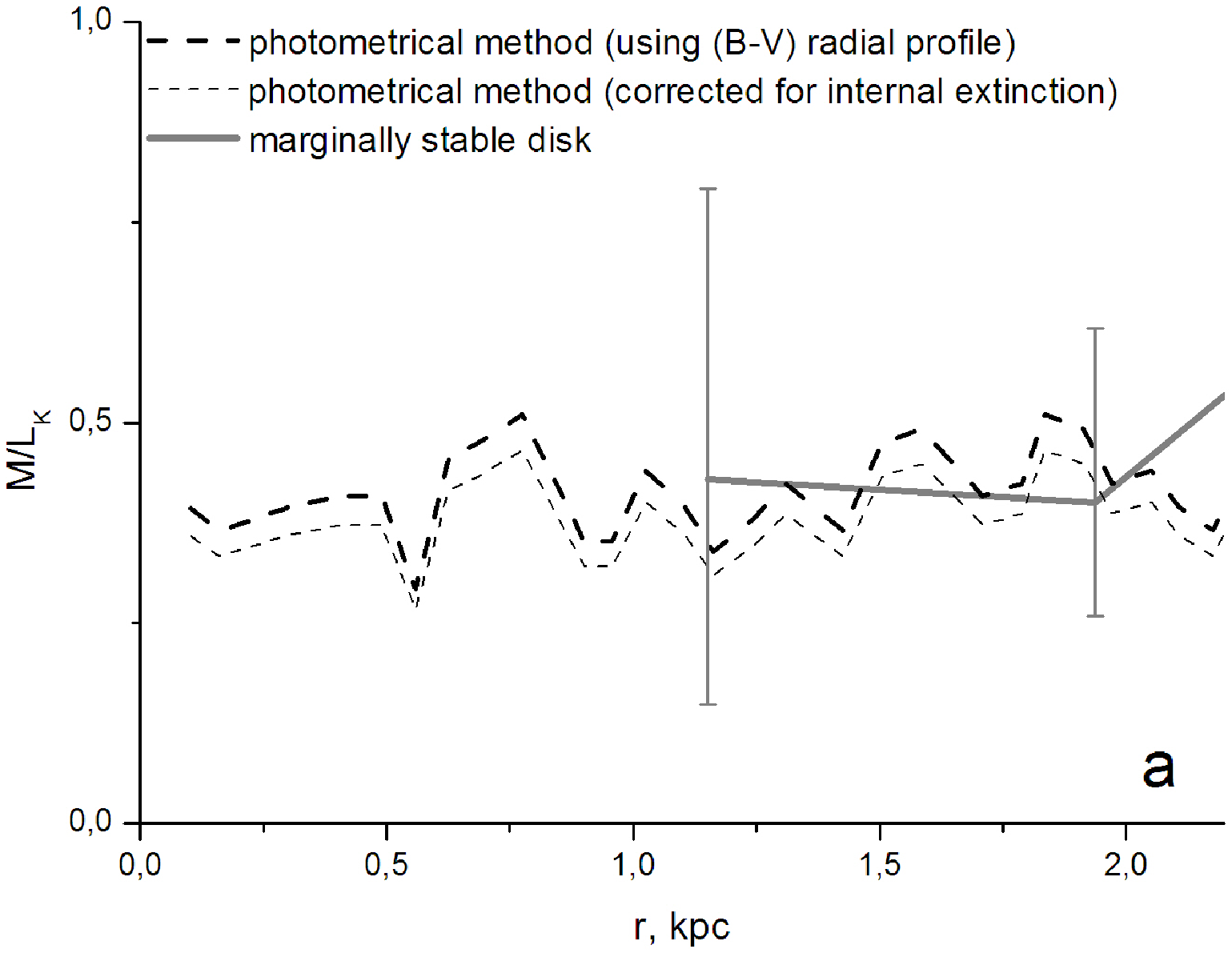}
\includegraphics[width=6.0cm,keepaspectratio]{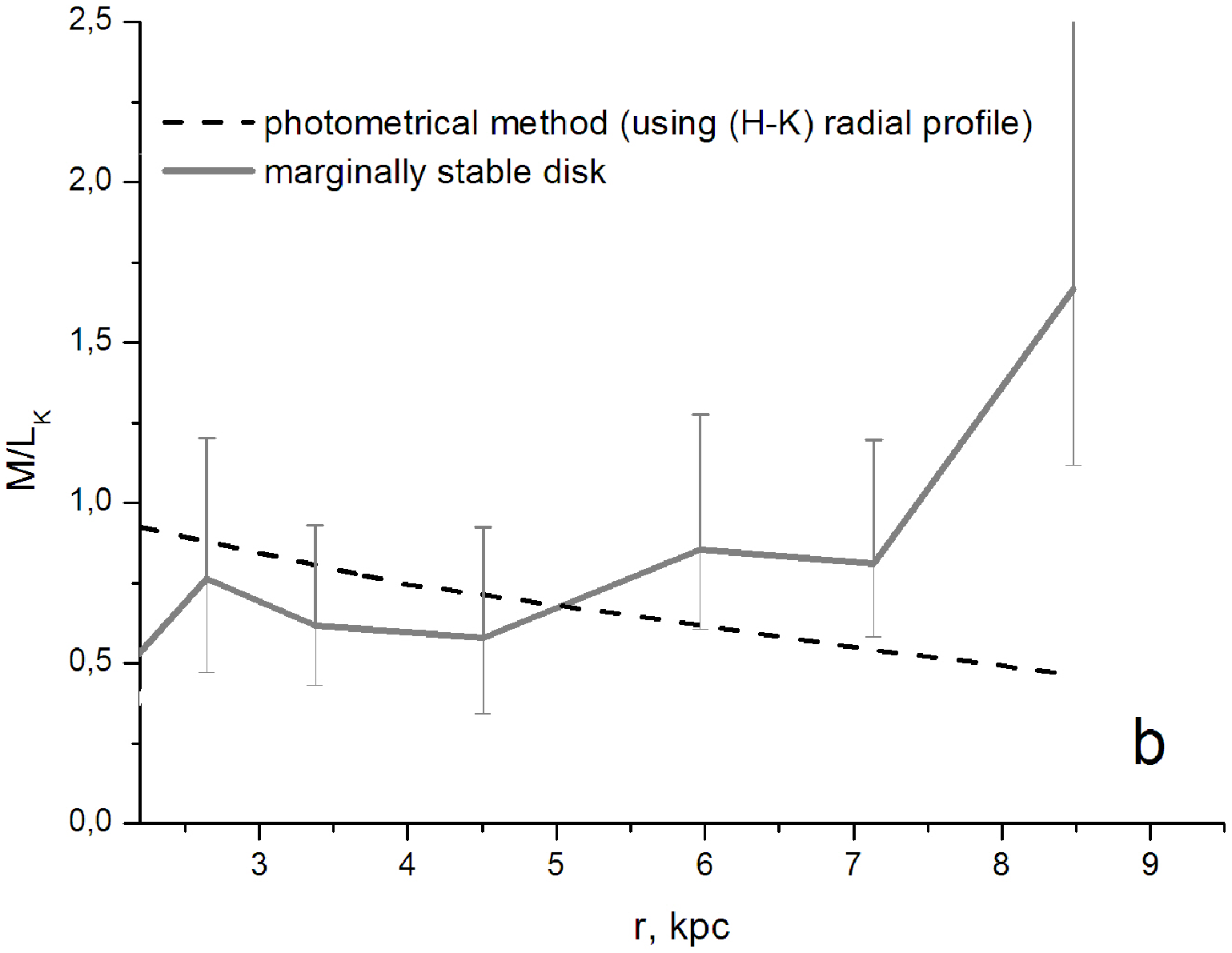}

\caption{Radial profiles of the mass-to-light ratio $M/L_K$ of M33 assuming that the
stellar disk is marginally stable and $m=c_z/c_r=0.4$ (solid line). The dashed lines correspond to the
profiles based on the color indices and stellar population models by Bell, de
Jong, 2001: (a) for the inner part of the galaxy and $(B-V)$ profile and (b) for the outer
disk and $(H-K)$ profile.}
\end{figure}
From Fig. 3 it follows that the surface density of the disk of M33, corresponding to the marginal
gravitational stability, is in a good agreement with the photometry-based
estimates --- with the exception of the most distant point ($r>7$ kpc), where the dynamical
overheating is quite possible. It  gives evidence that the disk of M33 within several radial scalelengths
have not
experienced a significant dynamical heating   or minor
merging events during its evolution.

The situation may be different for giant low surface brightness (LSB) galaxies. We applied the
gravitational stability criterion (1) to four LSB-galaxies, for which the distribution of
the velocity dispersion and rotation curves of ionized gas and stars, parallel with the
photometrical profiles in R-band, were given by Pizzella et al., 2008 (for more details
see Saburova 2011). The obtained surface density profiles were used to estimate $(M/L_R)_{d}$
for different galactocentric
distances (see Fig. 4). The filled and open symbols in Fig. 4 correspond to the estimates
based on the gas and stellar circular velocities. The dashed lines denote $(M/L_R)_{d}$
ratios, predicted by stellar population models of Bell, de Jong, 2001 for the metallicity
$Z=0.008$.
\begin{figure} [h!]
\vbox{
{\psfig{figure=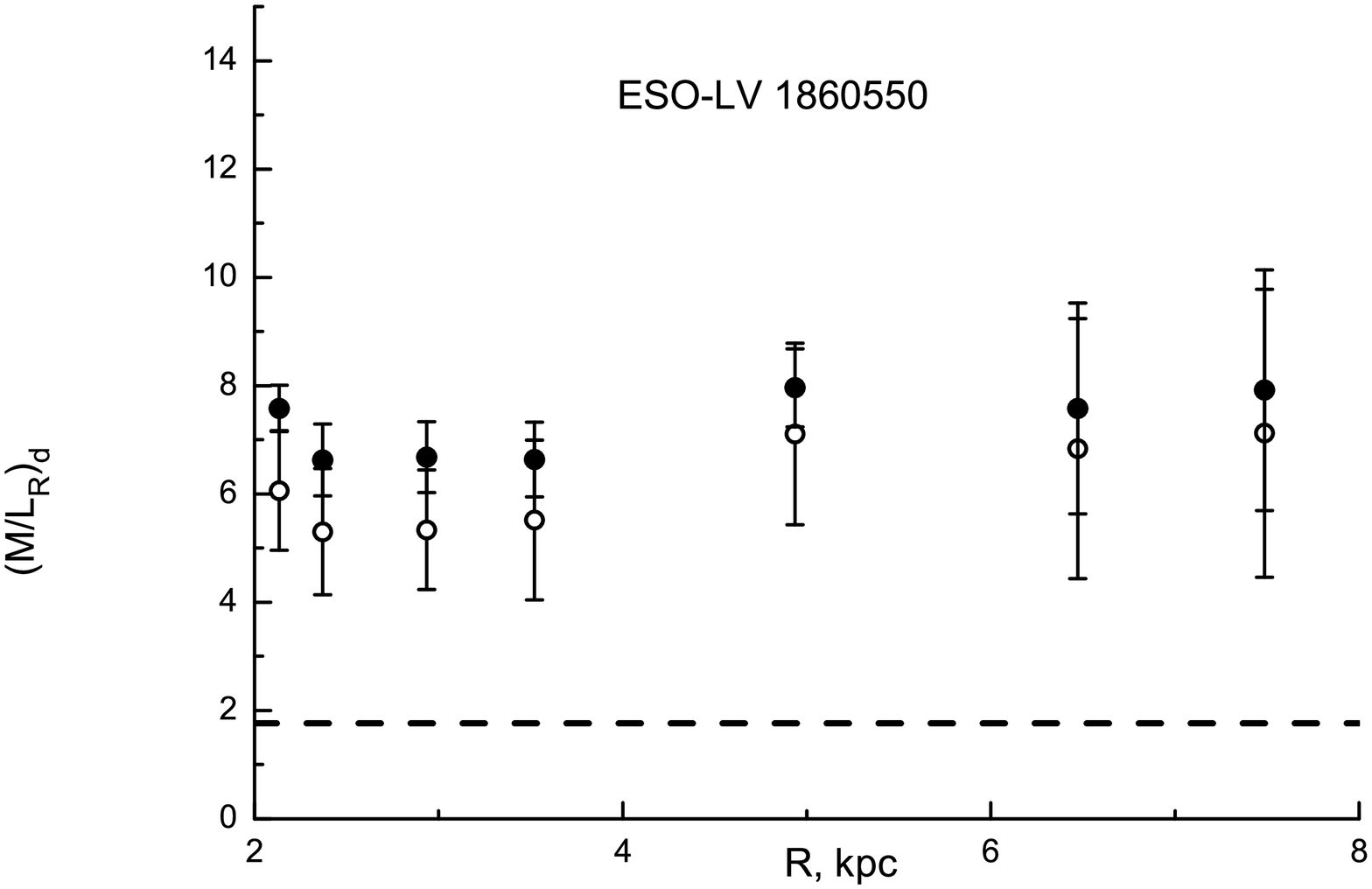,width=55mm,angle=0,clip=}}
\hspace{1mm}
{\psfig{figure=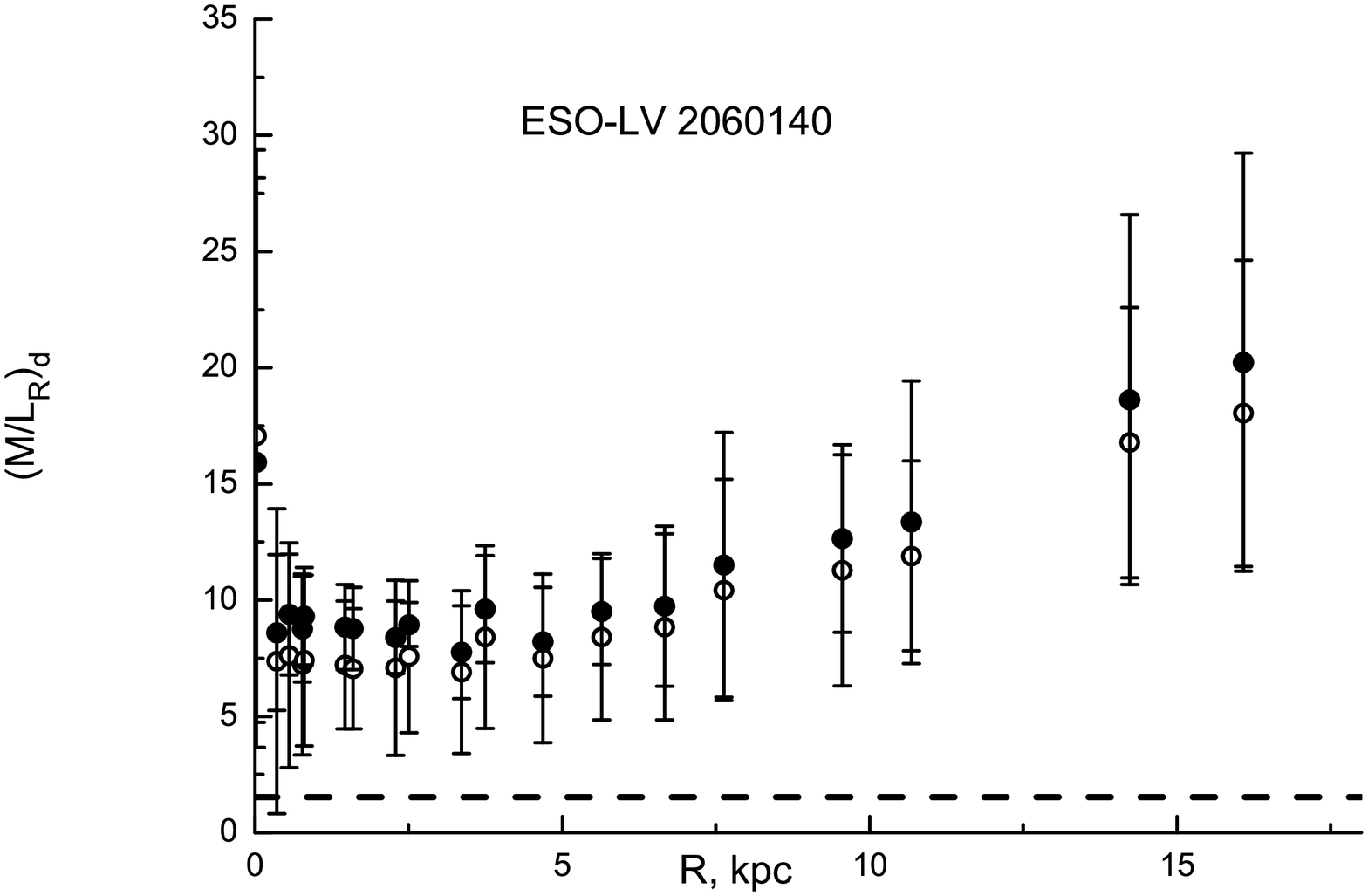,width=55mm,angle=0,clip=}}
\hspace{1mm}
{\psfig{figure=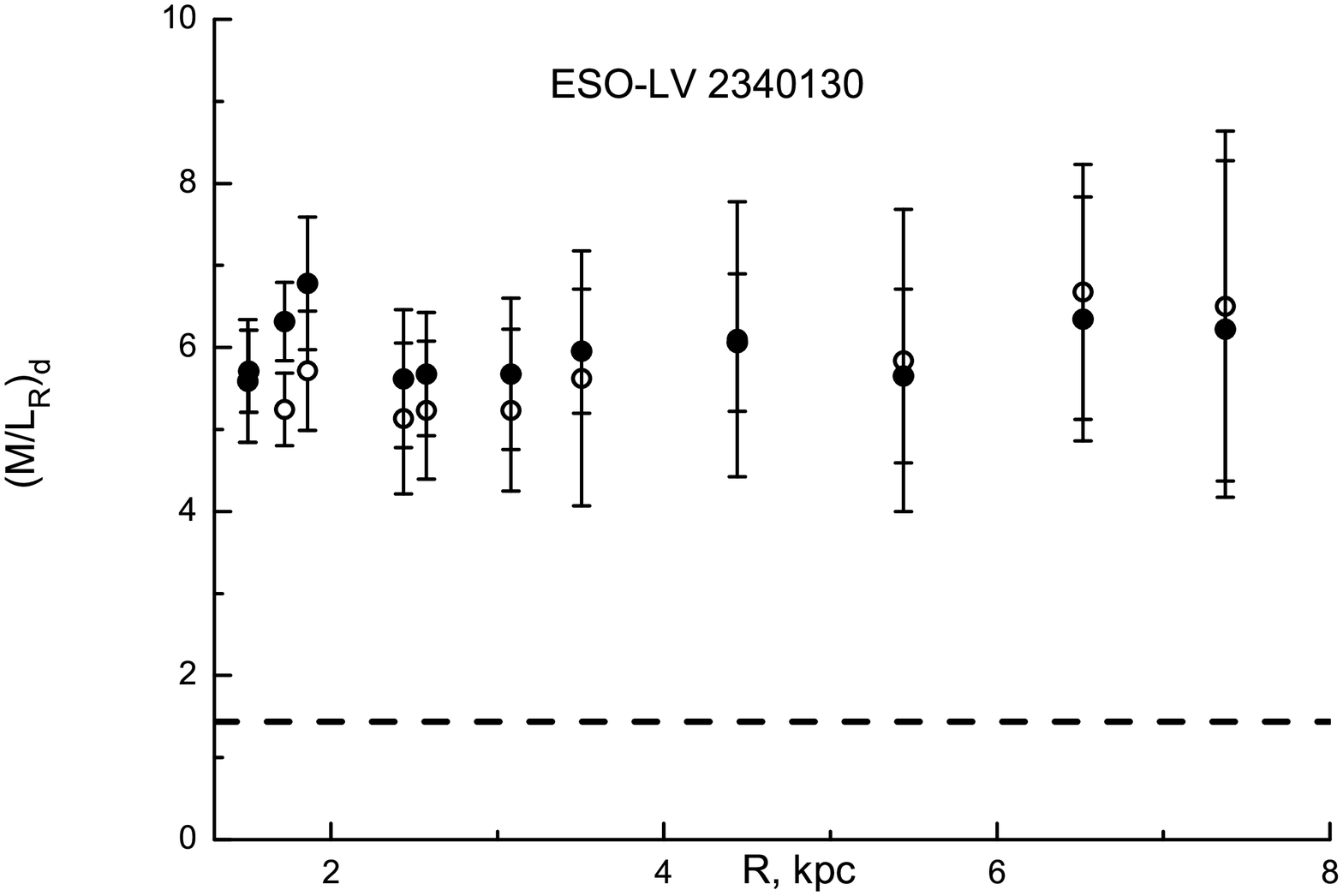,width=55mm,angle=0,clip=}}
\hspace{12mm}
{\psfig{figure=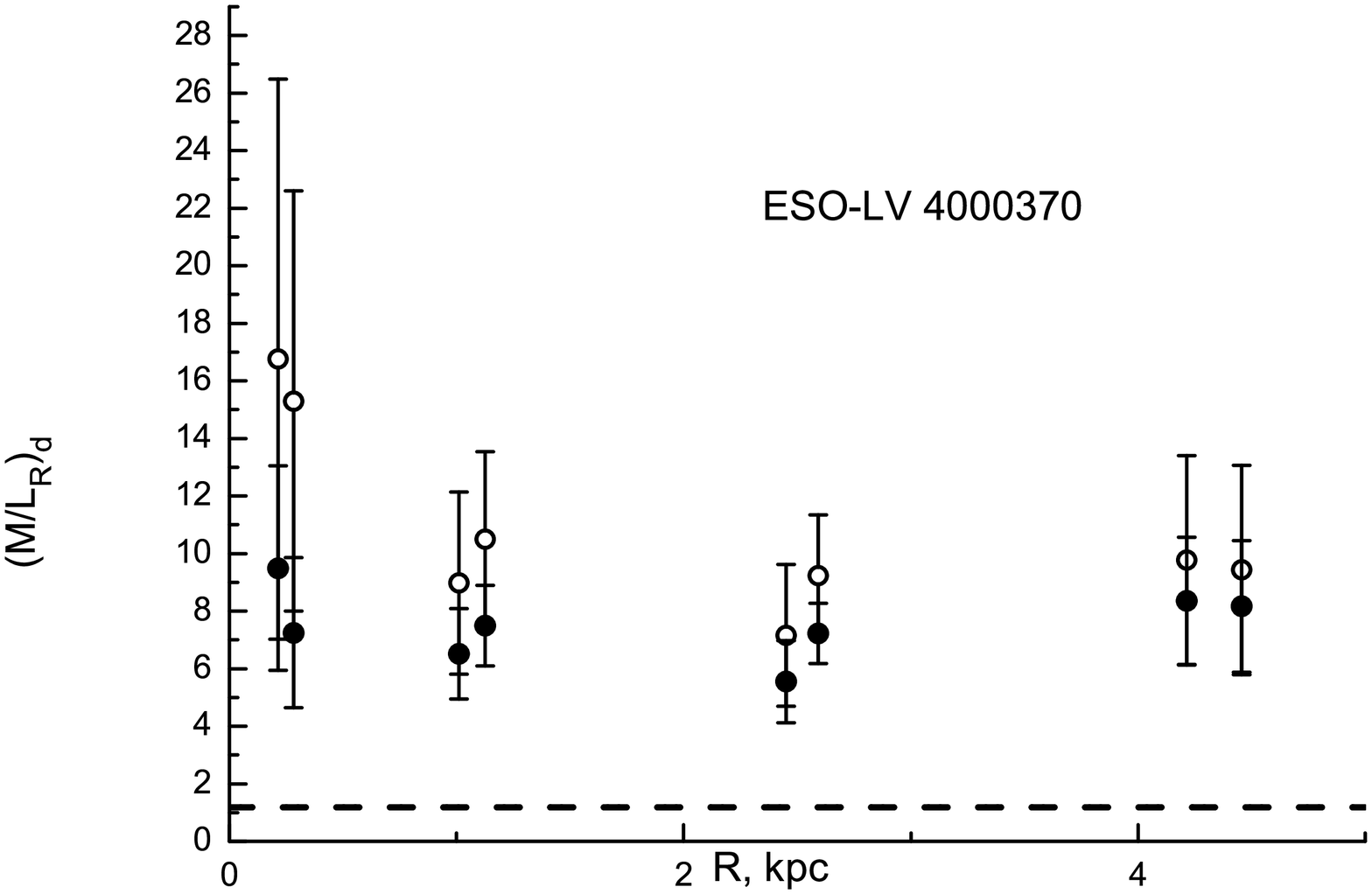,width=55mm,angle=0,clip=}}

\caption{Radial profiles of $(M/L_R)_{d}$ ratios for LSB-galaxies, obtained by applying  the local
gravitational stability criterion. Open and filled circles correspond to the data based on
the gas and stellar circular velocities. The dashed lines denote $(M/L_R)_{d}$ ratios,
predicted by stellar population models of Bell de Jong 2001.}}
\end{figure}
From Fig. 4 it follows that $(M/L_R)_{d}$ ratios which satisfy the  marginal stability
criterion are several times higher than those predicted by stellar population synthesis
models. It can indicate either a strong dynamical overheating of LSB disks, or the presence of a large
amount of dark (or faintly luminous) matter in disks. The latter case means, that the disks may
contain the unusually large amount of low massive stars (a heavy bottom IMF, see Lee et
al. 2004), or a significant fraction of non-stellar dark matter --- either in baryonic (e.g.
cold molecular gas, see Pfenniger, Combes, 1994) or non-baryonic form.

\sectionb{4}{CONCLUSION} 
Here we show on concrete examples for different types of galaxies that the comparison of  local velocity dispersion of stellar disk with the minimal value needed for its gravitationally stability enables to put  the  constraints on its  density and dynamical evolution.

\thanks{ This work was supported by Russian Foundation for Basic Research, grant 11-02-12247.}
\References
\refb Bell E. F., de Jong R. S. 2001, ApJ. 550, 212 


\refb Bottema R. 1993, A\&A, 275, 16


\refb Ciardullo R.,~Durrell P.R.,~Laychak M.B. et al. 2004 ApJ, 614, 167

\refb Corbelli E. 2003, MNRAS, 342, 199

\refb Guidoni U., Messi R., Natali G. 1981, A\&A, 96, 215

\refb Khoperskov A. V., Zasov ~A. V., Tyurina ~N. V. 2003, ARep, 47, 357

\refb Lee H., Gibson B.K., Flynn C. et al. 2004, MNRAS, 353, 113

\refb McGaugh ~S. S. 2005, ApJ, 632, 859

\refb Pfenniger D., Combes F. 1994, A\&A, 285, 94

\refb Pizzella A., Corsini E. M., Sarzi, M. 2008, MNRAS, 387, 1099

\refb Regan M.W., Vogel S.N. 1994, ApJ, 434, 536

\refb Saburova A.S. 2011, ARep, 55, 409

\refb Saburova \& Zasov 2012 in press

\refb Shapiro ~K.L., Gerssen J., van der Marel R.P. 2003, AJ, 126, 2707

\refb Verley S.,~Corbelli E.,~Giovanardi C.,~Hunt L.K. 2009, A\&A., 493, 453

\refb Zasov A.V., Moiseev ~A.V., Khoperskov ~A.V., Sidorova ~E. A. 2008, AstL, 52, 79

\refb Zasov A.V., Sil'chenko O.K. 2010, UFN, 180, 4

\refb Zasov A.V., Khoperskov A.V., Saburova A.S. 2011, AsL, 37, 374 

\end{document}